\documentclass[review]{elsarticle}

\usepackage{times,graphicx,subfig,amssymb}
\usepackage[linesnumbered,ruled,algonl,vlined,noend]{algorithm2e}
\newcommand{\var}[1]{\mbox{\emph{#1}}}
\newcommand{\myparagraph}[1]{\vspace{0.0pt}\noindent{\textbf{#1.}}~}

\newcommand{\ctq}{CTQ}
\newcommand{\qr}{QR-tree}
\newcommand{\qqr}{Q$^2$R-tree}

\journal{Journal of Big Data Research}

\begin{document}

\begin{frontmatter}
\title{An Efficient Index for Contact Tracing Query in a Large Spatio-Temporal Database}

\author[BUET]{Mohammed Eunus Ali}
\ead{eunus@cse.buet.ac.bd}

\author[BUET]{Shadman Saqib Eusuf}
\ead{s.saqibeusuf@gmail.com}

\author[UVA]{Kazi Ashik Islam}
\ead{ikaziashik@gmail.com}

\address[BUET]{Bangladesh University of Engineering and Technology, Bangladesh}
\address[UVA]{University of Virginia, USA}

\begin{abstract}
In this paper, we study a novel contact tracing query (CTQ) that finds users who have been in \emph{direct contact} with the query user or \emph{in contact with the already contacted users} in subsequent timestamps from a large spatio-temporal database. The CTQ is of paramount importance in the era of new COVID-19 pandemic world for finding possible list of potential COVID-19 exposed patients. A straightforward way to answer the CTQ is using traditional spatio-temporal indexes. However, these indexes cannot serve the purpose as each user covers a large area  within the time-span of potential disease spreading and thus they can hardly use efficient pruning techniques. We propose a multi-level index, namely {\qr}, that consider both space coverage and the co-visiting patterns to group users so that users who are likely to meet the query user are grouped together. More specifically, we use a quadtree to partition user movement traces w.r.t. space and time, and then exploit these space-time mapping of user traces to group users using an R-tree. The {\qr} facilitates efficient pruning and enables accessing only potential sets of user who can be the candidate answers for the CTQ. Experiments with real datasets show the effectiveness of our approach.
\end{abstract}

\end{frontmatter}

\section{Introduction}
The world is witnessing an unprecedented pandemic as the coronavirus (SARS-CoV-2) is continuing its spread across the globe. As of today (May 13, 2019) there are more than four million confirmed cases in 185 countries and more than 290,000 people have lost their lives due to the virus infected respiratory infection commonly referred to as COVID-19~\cite{coronaJHU}. This virus is extremely infectious, where it can easily pass from person to person. Thus, to curb the spread of the coronavirus, authorities around the world implemented lockdown measures for months. However, these lockdowns have brought much of global economic and social activity to a halt. 

To avoid the socio-economic catastrophes, the authorities have gradually started to ease the lockdowns. However, they are still struggling to find efficient techniques to monitor the mobility of potentially COVID-19 infected patients and who have been in contact with a virus infected person. Since people in close contact with someone who is infected with the virus are at higher risk of becoming infected themselves, and of potentially further infecting others, closely monitoring these contacts can prevent further transmission of the virus. This process of monitoring is known as \emph{contact tracing}. In this paper, we study the problem of contact tracing query (CTQ) in a spatio-temporal database. 

Consider the following scenario. Let $D$ be the historical mobility traces (or equivalently trajectories) of users for the last $T$ days, obtained from GPS-enabled phones or mobile signals through triangulations. Thus, each user $u \in D$ is represented as a sequence of time stamped locations $\{ (l_1, t_1), (l_2, t_2)....(l_n,t_n)\}$ denoting her visited places $l_1, l_2, ..., l_n$ in different times $t_1, t_2, ..., t_n$, respectively. Let $q$ be the mobility traces (or the trajectory) of a newly identified COVID-19 infected user, which is the query in our system. The objective of the CTQ is to identify a set of users $U \subset D$  who have been in direct contact with $q$ at any point of time, and subsequently find users who came into contact with the already contacted users.


To process variants of trajectory related queries such as range, join, nearest-neighbor, etc., a large body of trajectory indexing techniques have been proposed in the literature~\cite{Mokbel2009, Ali18, Shang2017, chakka2003indexing}. These indexes are variants of traditional spatio-temporal indexes such as $R$-tree~\cite{BeckmannR} or quad-tree~\cite{Hanan84}. These indexes are tailored for answering different types of queries. Though it may seem that the CTQ can be solved by using existing indexes designed for range queries, running repetitive range queries for different points of the query trajectory in the CTQ will make it extremely in-efficient. This is due to the  following two reasons: (i) the mobility traces or historical trajectories of a user  are usually a set of time-stamped dispersed point locations covering a large area, which is different than the normal point data such as POIs (Point of Interest) or trajectory data such as taxi trips, and thus very hard to prune using traditional indexes, (ii) if a user's travel history matches with the query at any instance then the user will be a candidate answer, and we need to run the process recursively as this user may have subsequently infected others.


To answer the {\ctq} efficiently, we propose a two-level index structure, namely $QR$-tree, that exploits the strengths of both quad-tree and R-tree. In the first level, we use a quad-tree to partition the points of historical trajectories, where the location of a trajectory point is specified by the $space-id$ of the smallest quad-tree block that contains the point. Similarly, the timestamp of each location of a trajectory is mapped to a $time-id$ that corresponds to a time bucket containing the timestamp of the trajectory point. After that, we transform each trajectory as a sequence of $(space-id, time-id)$ tuples. We consider this mapping  as a transformation to a new coordinate system for the trajectory points. Next, we apply an R-tree on the trajectory points, represented by the new coordinate system, for grouping and saving them in disk. Finally, we present an efficient divide-and-conquer approach to answer {\ctq}, where a query is recursively divided and run through different levels of the index to find the users who match in both space and time.


The contributions of the paper are summarized as follows:

\begin{itemize}
\item We are the first to introduce a novel contact tracing query ({\ctq}) in a large spatio-temporal database, which is of paramount importance for identifying users who were potentially exposed to COVID-19 infected users.
\item We propose a multi-level index structure, namely {\qr}, that combines the regular space-partitioning strategy of a quadtree, and the object grouping strategy of an R-tree, to organize the user spatio-temporal data in such a way that facilitates faster processing of the {\ctq}.
\item We present an efficient divide-and-conquer approach for answering {\ctq} queries using the {\qr}. We evaluate our indexes and algorithms through an extensive experimental study on real datasets, which demonstrate both the efficiency and effectiveness of our solution.
\end{itemize}

\section{Problem Formulation}
\label{problem}
Let $D$ be the historical mobility traces (or equivalently trajectories) of users for the last $T$ days, obtained from GPS-enabled phones or mobile signals through triangulation. Each user $u \in D$ is represented as a sequence of time stamped locations $\{(l_1,t_1), (l_2, t_2), ... (l_n,t_n)\}$ denoting her visited places $l_1, l_2, ..., l_n$ in different times $t_1, t_2, ..., t_n$, respectively. Let $q$ be the mobility traces (or the trajectory) of a COVID-19 infected user, which is represented as the sequence $\{ (q.l_1, q.t_1), (q.l_2, q.t_2)....(q.l_n, q.t_n)\}$, the query in our system. 

Any two users $u$ and $v$ \emph{meet} each other if and only if both the spatial distance and temporal distance of any two points of $u$ and $v$, respectively, are under certain distance thresholds. Formally, the meeting condition for two trajectories can be expressed as follows.

\emph{Condition of $u$ meets $v$: } Let $\{ (u.l_1, u.t_1), (u.l_2, u.t_2)....(u.l_n, u.t_n)\}$ and $\{ (v.l_1, v.t_1),$ $(v.l_2, v.t_2)....(v.l_m, v.t_m)\}$  be the two sequence of time-stamped locations of $u$ and $v$, respectively. Let $spatialDist$ and $temporalDist$ be the spatial distance and temporal distance measuring functions between two locations and two time-stamps, respectively. Now, for any $i \in [1, n], j \in [1, m]$, if $spatialDist (u.l_i, v.l_j) \leq \psi$ and $temporalDist (u.t_i, v.t_j) \leq \tau$, then we say the trajectory $u$ \emph{meets} the trajectory $v$. Here, $\psi$ and $\tau$, are spatial (euclidean) and temporal distance thresholds, respectively.


The objective of the CTQ is to identify the set of users $U \subset D$ where each user $u \in U$ has potentially been \emph{exposed} to the corona virus. We define the set $U$ as follows.
\begin{enumerate}
\item Let $U_0$ be the set of users where each user $u \in U_0$ met with $q$ at any timestamp $u.t$ in the last $T$ days. We say that $u$ was exposed at time $u.t_{exposed}\ (=u.t)$. Here, the subscript zero (\emph{0}) in $U_0$ denotes that, the number of intermediate carriers (of the virus) between user $q$ and $u$ is zero (i.e $u$ was infected by the query user $q$).
\item Let $U_1$ be the set of users where each user $u \in U_1$ met with any user $v \in U_0$ at timestamp $u.t > v.t_{exposed}$. We define $u.t_{exposed}\ (=u.t)$ as the time of exposure for user $u$.
\item In a similar manner, we can define the set of users $U_i$ recursively where each user $u$ was exposed to some user $v \in U_{i-1}$. Then, we define the set $U$ as: $$U =  \bigcup\limits_{i=0}^{L-1} U_i$$ Here, $L$ is an integer that denotes the maximum allowed depth for the recursion and is passed as a parameter for \ctq.
\end{enumerate}




Based on the above definitions, we formally define our contact tracing query as follows.

\begin{description}
\item[\emph{Definition 2.1.} \ctq.]
Given a set $D$ of user trajectories, a COVID-19 infected user trajectory $q$, a spatial proximity threshold $\psi$, a temporal proximity threshold $\tau$,  and an integer $L$, a {\ctq} query finds a set of users $U \subset D$ such that $U = \bigcup_{i=0}^{L-1} U_i$; where $U_i$ is the set of users who were \emph{exposed} to the query user through `$i$' number of intermediate carrier users.
\end{description}

\section{The Proposed Index}
\label{index}
The trajectories in our datasets can be very long (e.g., last 14 days of mobility traces of each user) and may cover large areas. Using an R-tree to index the two spatial and one temporal dimension of these trajectories might not be useful as each trajectory's MBR (Minimum Bounding Rectangle) will most likely overlap with too many other trajectories' MBRs, making the pruning scheme of the R-tree ineffective. On the other hand, if we use a quadtree to index all points of the trajectories, the points of a single trajectory may end up in many quadrant of the quadtree blocks, and thereby making it hard to decide which of those trajectories should be stored together in a disk block to facilitate faster retrieval of candidate users.

\subsection{\qr}
The key intuition of our proposed index is, the trajectories whose points are co-located at the overlapping time-instant are likely to match with the same query. Based on this observation, we present a two-level index, the Quad R (QR) tree, that combines the strengths of both quadtree and R-tree.

First, the spatial data space is recursively partitioned using quadtree, where each leaf quadtree block does not contain more than $\theta$ points. Then we use a space filling curve, specifically a z-curve (Morton order), to number these leaf quadtree blocks. We call such a number, the $spatial-id$ of the block. Thus, in the spatial domain each trajectory is represented as a list of  $spatial-id$s. Similarly, each timestamp of a trajectory is mapped to a time bucket and assigned a number $temporal-id$. Thereby, each trajectory can now be represented as sequence of ($spatial-id, temporal-id)$ tuples. This new mapping of trajectories can be seen as a transformation to a new coordinate system, where $x$-axis represents spatial dimension and $y$-axis represent temporal dimension, and each trajectory is represented as a set of points in that space.

In the next step, we use an R-tree to group trajectories based on their sets of points in the transformed space. Essentially, each set of points in this new space is represented as an MBR, and the R-tree groups close-by MBRs in a leaf node. Each leaf node of the constructed R-tree is stored in a disk-page. We maintain this \emph{disk-page id} in all the corresponding leaf-blocks of the first level quadtree that contain a point of the trajectories stored in this disk-page. In the quadtree block, we also maintain associated temporal ids denoting the time range of trajectory points stored in the corresponding disk-page. Note that, we do not keep the hierarchical structure of R-tree for query processing, rather we only use the R-tree for grouping of similar trajectories in the transformed space.

\begin{figure}
	\hskip-0.18cm\begin{tabular}{ c c }
        \includegraphics[width=1.8in, height=1.4in]{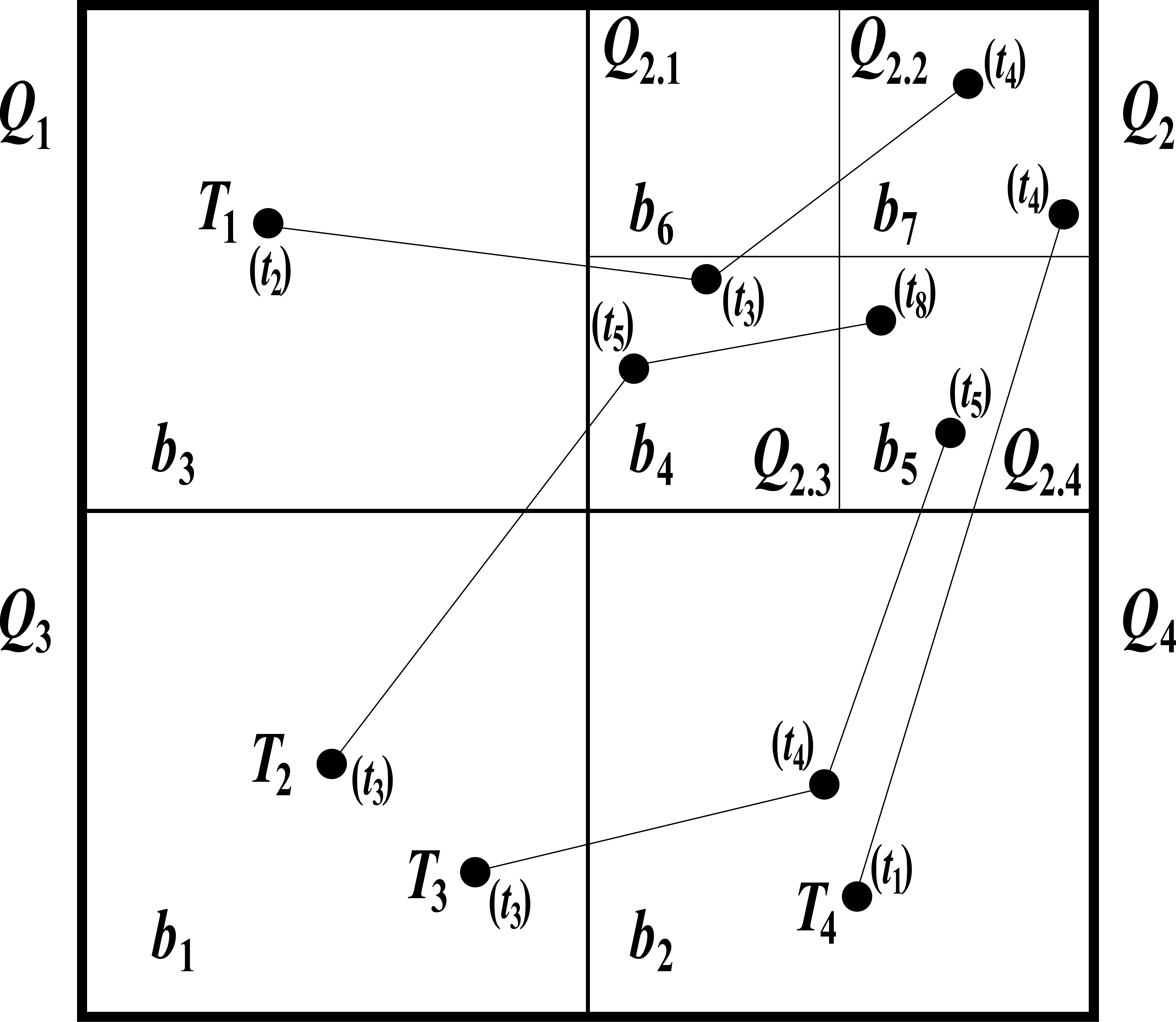} &
        \raisebox{0cm}{\includegraphics[width=1.6in,height=1.0in]{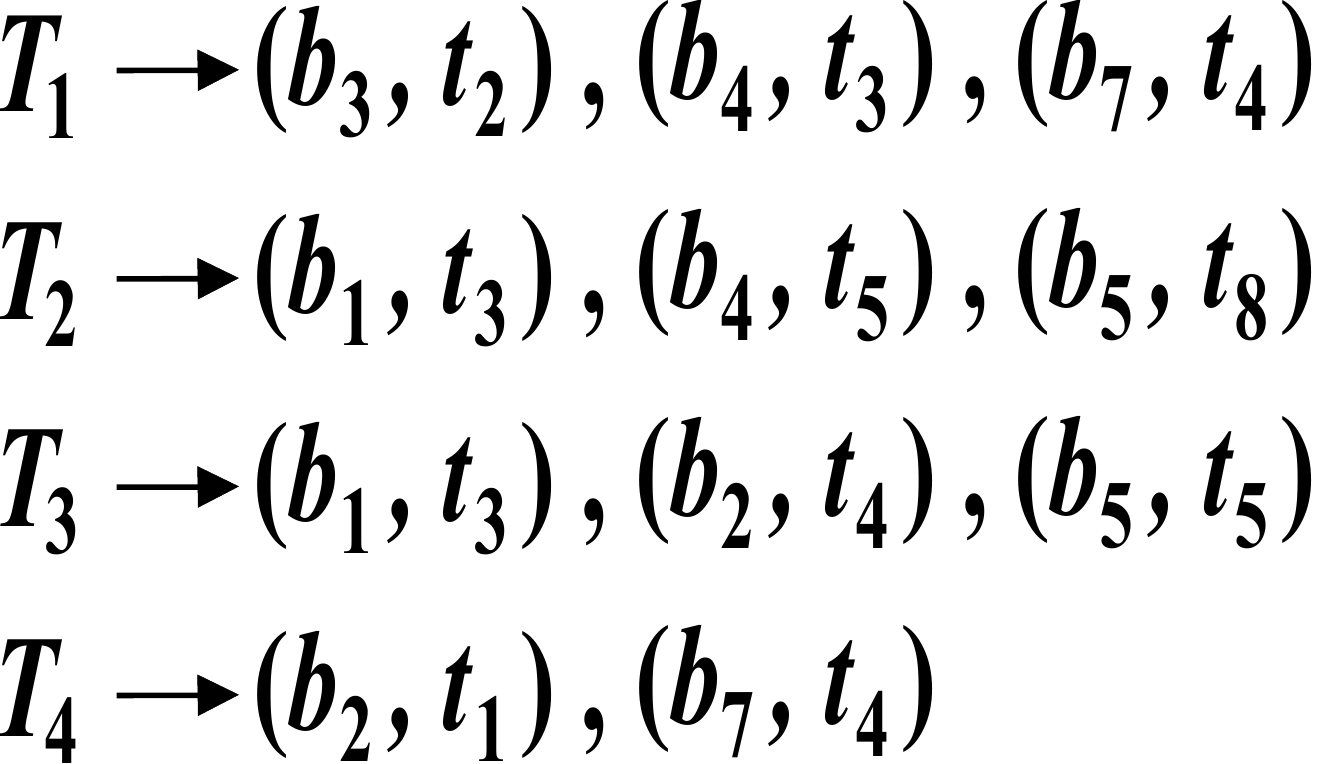}}\\
	\end{tabular}
    \caption{ (a) A quadtree based space partitioning of  trajectories. (b) Mapping of trajectories.}
    \vspace{-5pt}
    \label{fig:qrtree}
\end{figure}

\begin{figure}
	\hskip-0.18cm\begin{tabular}{ c c }
        \includegraphics[width=1.6in, height=1.4in]{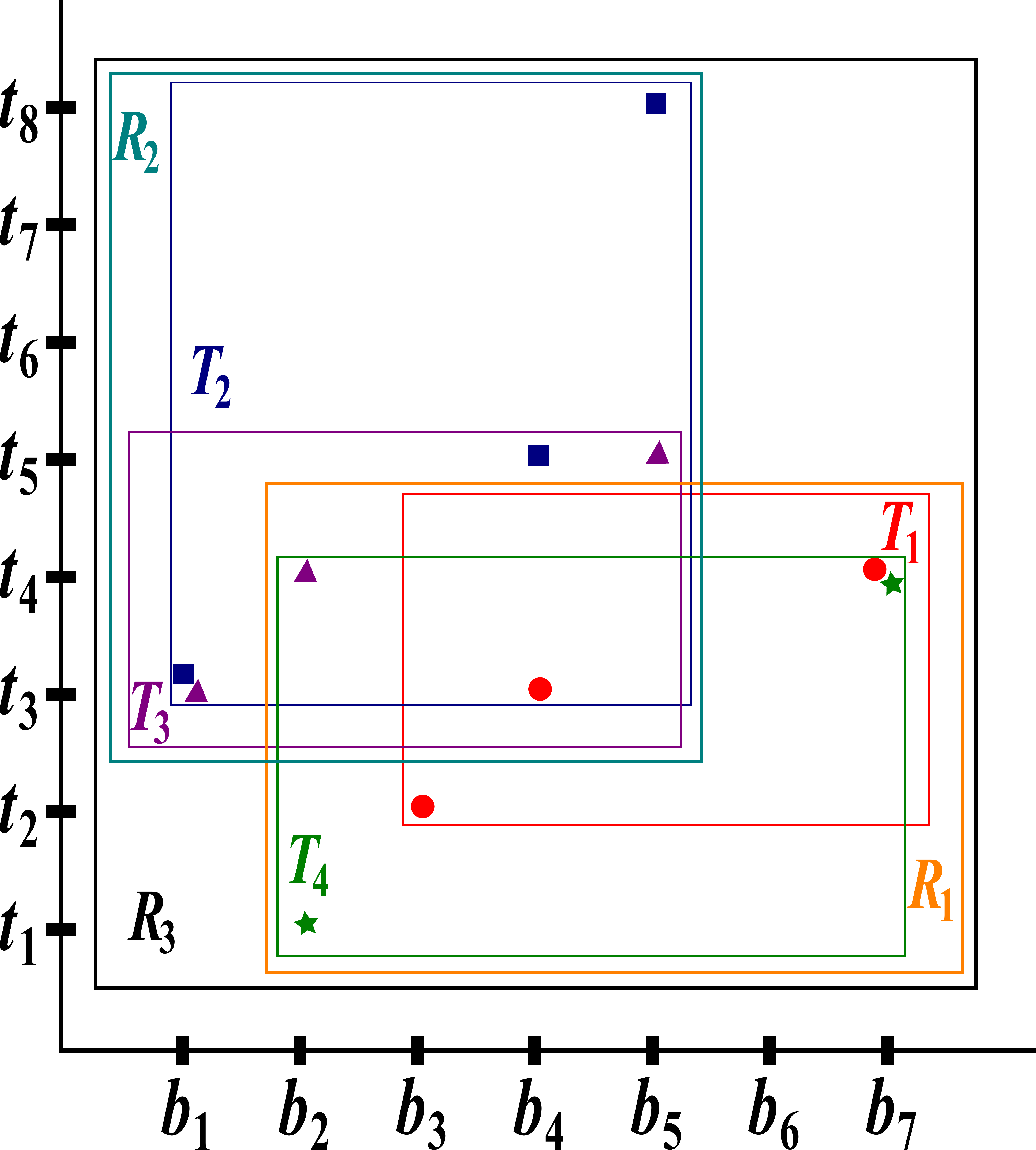} &
        \raisebox{0cm}{\includegraphics[height=1.4in]{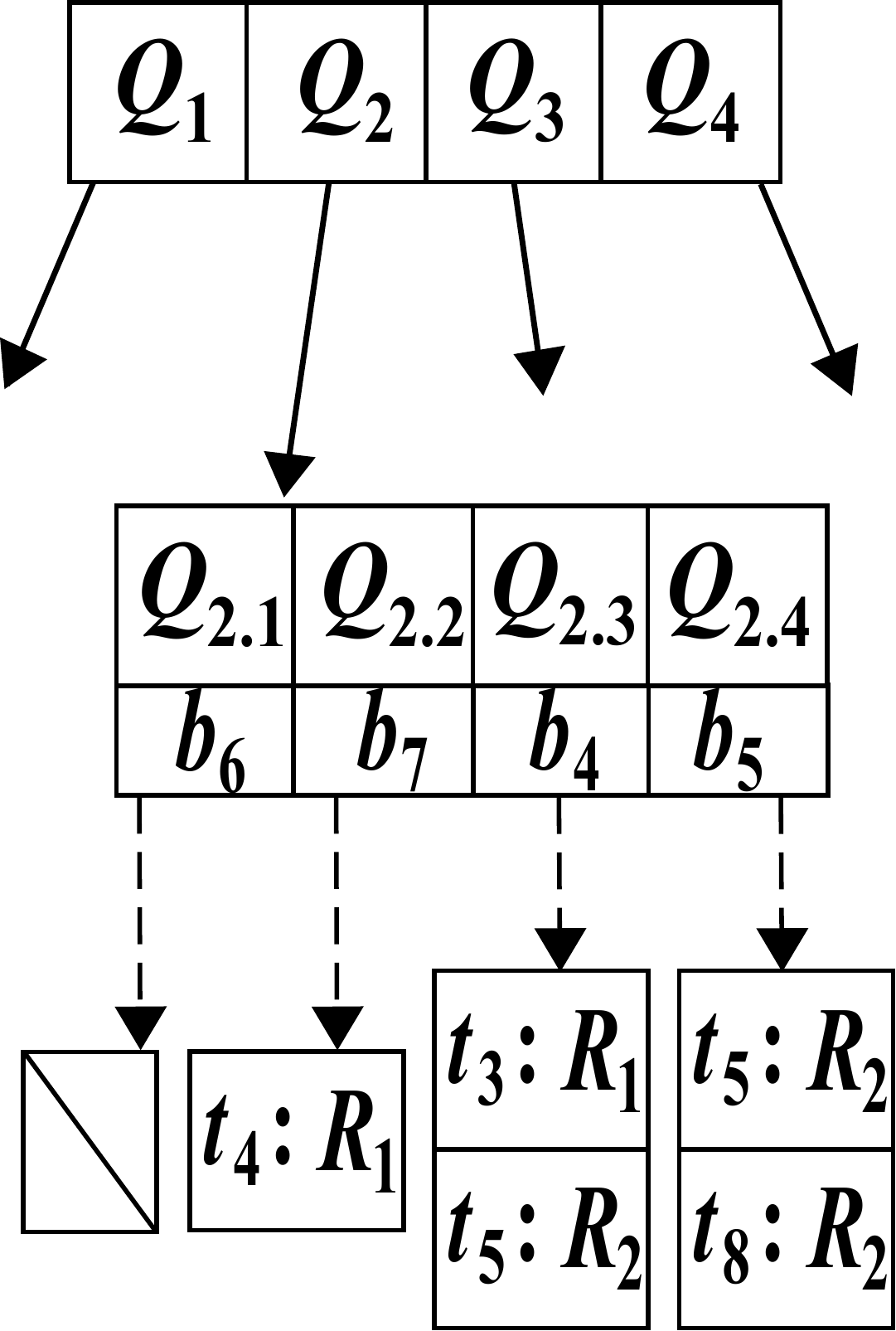}}\\
	\end{tabular}
    \caption{(a) An R-tree based grouping of trajectory points in a transformed space. (b) A QR-tree index structure.}
    \vspace{-8pt}
    \label{fig:qrtree1}
\end{figure}

{\textit {
Figure~\ref{fig:qrtree} and  Figure~\ref{fig:qrtree1} show the construction process of the QR-tree.  Figure~\ref{fig:qrtree}(a) shows an example with four user trajectories, $\{u_{1}, \dots, u_{4}\}$, where $\theta = 2$. The space is first divided into four quadrants $Q_1, \dots,Q_4$.
As $Q_2$ contains more than $2$ trajectory points, this block is further divided into  $Q_{2.1}, \dots,Q_{2.4}$. We then apply z-ordering to number these quadtree blocks as $b_1, b_2, ..., b_7$. After that each time-stamp of points in the trajectories is assigned time-bucket number between $t_1$ and $t_8$. After that  Figure~\ref{fig:qrtree}(b) shows the new representation of trajectories $u_1- u_4$ as a sequence of $(b_i, t_j)$ tuples. These points are then mapped into a new co-ordinate system in a two-dimensional space (Figure~\ref{fig:qrtree1}(a)), where we can see points of four trajectories in four different colors, and each set of points of a single trajectory is represented as an MBR. These MBRs are grouped together to form an R-tree. Each leaf level node, $R_1, R_2$ corresponds to a disk-page.  Finally, we maintain these disk-page references in different level quadtree blocks of the QR-tree, as shown in (Figure~\ref{fig:qrtree1}(b)). For example, with $Q_{2.4} (b_5)$, disk-page id $R_2$ is assigned along with time-bucket ids $t_5$ and $t_8$.}}

\subsection{\qqr}

We make further improvement on the proposed {\qr} index, where we augment the  index by adding another top-level quadtree. The intuition behind adding this top level quadtree is two fold: (i) it partitions the entire sets of trajectories into different groups based on their extents, thus the index will have better pruning capability, (ii) since a longer trajectory will most likely contain more points than a shorter trajectory, maintaining different length trajecotory in a single R-tree is challenge as trajectories of different length may occupy different storage spaces in the disk.

In our proposed \qqr, a trajectory is stored under any non-leaf or leaf nodes based on their extents. In this case, we recursively partition the space, and a trajectory is stored in a quadtree block that fully contains it. Thus, long trajectories are stored in the upper level quadtree blocks than shorter trajectories are stored in lower level trajectory blocks. For all the trajectories in a single quadtree block, we apply {\qr} strategy to organize them in disk. Since we use two quadtrees and one R-tree in the index, we refer this index as {\qqr}.
\section{Processing \ctq}
\label{algorithm}

In this section, we present an algorithm for processing {\ctq} using our proposed \qr. Intuitively, the quadtree of {\qr} supports faster range query around query trajectory points, while R-tree grouping ensures the lower I/O overhead. We apply a spatial pruning followed by a temporal pruning using \qr, where the irrelevant quadtree nodes are pruned first, and then the time buckets are used to further prune the R-tree blocks to be retrieved. For simplicity, we present the first level contact tracing, where the task is to find users who were directly exposed to $q$. 

\subsection{{\ctq} Matching Algorithm}
\setlength{\algomargin}{1.2em}
\begin{algorithm}[t]
    \caption{matchCT($N$,$q$)}
    \label{alg:match_ct}
    \begin{smaller}
    \KwIn{A quadtree node $N$ of {\qr}, a COVID-19 positive user trajectory $q$}
    \KwOut{A set $U$ of user trajectories suspected to be exposed by $q$}
    $U \gets \varnothing$ \\
    \lIf {$q = \varnothing$} {\Return U\\} \label{a1_end1}
    \uIf {$N$ is a $leaf$} {
		$t_{b\_list} \gets$ extendedTimeWindows($q$)\\
		$U \gets$ evaluateContacts($N, t_{b\_list} , q$) \label{a1_end2}
	}
    $N_{children} \gets$ children($N$) \\
    $q_{children} \gets$ extendedIntersection($N_{children},q$)\\ \label{line:a1_fi} 
    \For{$N_c \in N_{children}$, $q_c \in q_{children}$\label{line:a1_for} } 
    {
    	$U \gets U \ \bigcup$ matchCT($N_c, q_c$) \label{line:a1_recur}
    }
    \Return $U$
    \end{smaller}
    
\end{algorithm}

Algorithm~\ref{alg:match_ct} describes the pseudocode for a divide-and-conquer algorithm for the \ctq. A user $u$ can be infected by $q$, if a point of $u$ is within a threshold distance $\psi$ and a threshold time $\tau$ from any point of $q$. So to facilitate this spatio-temporal range search, we consider an extended minimum bounding rectangle (EMBR) (in terms of space) of every points of $q$ to include the  infectious region of $q$.  

Initially, the function $\var{matchCT}(\cdot)$ is called with the root node $N$ of the {\qr} and $q$. It finds the relevant child nodes of $N$ that intersect with $q$ (or EMBRs of $q$) in the function $\var{extendedIntersection}(\cdot)$ (Line~\ref{line:a1_fi}). Thus, quadtree nodes that are within $\psi$ spatial distance threshold from $q$ are considered. If a child node does not intersect with the EMBR, it can be safely pruned. Otherwise, each unpruned child node $N_c$ of $N$ and the corresponding components of $q$, are passed to $\var{matchCT}$ function according to Algorithm~\ref{alg:match_ct} (Line~\ref{line:a1_recur}).
 
The recursive method has two base conditions: (i) when $q$ is empty (there is no point left in that subspace for repeated division (Line~\ref{a1_end1})); and (ii) when $N$ is a leaf node. In case of a leaf node, the possibly infectious time buckets corresponding to the points in $q$ are calculated with the function $\var{extendedTimeWindows}(\cdot)$. This function returns all the possible time windows within temporal range $\tau$ starting from that of each point of $q$. Then the exposure of the trajectories stored in the disk blocks mapped to the node $N$ and entries of temporal bucket are computed with $\var{evaluateContacts}(\cdot)$.

The function $\var{evaluateContacts}$ is used to determine which trajectories meet with $q$. To compute it, first we need to retrieve trajectories which have transformed coordinates ($N, t$), for each entry $t \in t_{b\_list}$. So we look up our in memory {\qr} index and obtain a list of relevant $R-tree$ nodes (i.e. disk block ids). We the fetch the trajectories stored in those disk blocks. For each user trajectory $t_i \in T_r$, we compute whether the user meets with $q$.  The trajectory $t_i$ is included in the exposed set $U$  if it meets with $q$.

The above algorithm supports contact tracing by passing the depth level $L$ in $\var{matchCT}(\cdot)$ as a recursion depth parameter. Initially the parameter is set to 0. Then in the aforementioned second base condition, we can call $\var{matchCT}(\cdot)$ recursively for each of the computed exposed trajectories with depth parameter incremented by 1, until it has reached $L$. 
\section{Experimental Evaluation}
\label{experiment}
In this section, we compare the {\qr} with a baseline (\textbf{BL}) approach, where we use a 3D R-tree (for location and timestamp) for indexing. We use it as our baseline because, in contrast to other methods, a trajectory is saved in an Rtree leaf as a single object. This is ideal for retrieving the whole trajectory during the processing of {\ctq}. We use the mobility traces from the CDR data collected by Grameenphone Ltd  between June 19, 2012 and July 18, 2012~\cite{HasanA17}, hereafter referred to as \emph{BD Cellphone}, as our default dataset. Besides, we use Foursquare check-in dataset ~\cite{yang2014modeling} of New York city, hereafter referred to as \emph{NYF}, to evaluate performance of {\ctq} in a different spatio-temporal domain. We use JDK 1.8 for implementing our algorithms, which were run in Intel core i5-3570K processor (3.4 GHz) and 8 GB of RAM.

\myparagraph{Performance Evaluation and Parameterization}
The parameters we varied, their ranges and default values (in bold) are shown in Table~{\ref{tab:param}}. We have varied a single parameter in each experiment while the others are assigned their default values. We measure the impact of the parameters on runtime and I/O cost i.e. \# of disk blocks accessed in {\ctq} processing. We configure the quadtree nodes to hold upto 128 points, and R-tree blocks to hold upto 4 trajectories. For each set of experiments, we run 100 {\ctq} and present the average result.

\begin{table}[h]
	\begin{smaller}
		\begin{center}
			\begin{tabular}{ |l|l| }
				\hline
				Parameters & Ranges \\
				\hline 
				\# of Points Per Query Trajectory & 1-50, {\bf 51-100}, 101-200, $>$ 200\\
				\# of Trajectories Indexed & 10k, 25k, {\bf 50k}, 100k \\
				Spatial Distance Threshold ($\psi$) & 1m, {\bf 2m}, 4m, 10m\\
				Temporal Distance Threshold ($\tau$) & 1 min, 15 min, {\bf 30 min}, 1 hour, 3 hour\\
				Maximum Recursion Depth ($L$) & {\bf 1}, 2, 3\\
				\hline
			\end{tabular}
		\end{center}
	\end{smaller}
	\vspace{-1pt}
	\caption{Parameters}
	\label{tab:param}
\end{table}

Note that, the choice of spatial and temporal range thresholds is mostly application specific. We have varied them in the aforementioned range mainly to demonstrate the performance of our work. For COVID-19, 1 meter was considered as the maximum distance for transmission via respiratory droplets \cite{whocovid19}, which is suggested as 2 meters in some other studies\cite{tbsnews}. Besides, there is an evidence, but perhaps no concrete proof of transmission by aerosolized respiratory fluids, which is in fact likely to travel farther. So we have varied the spatial range upto 10 meters, considering situations of indoor environment. On the other hand, there is still no authentic information on the temporal threshold for COVID-19 transmission. To the best of our knowledge, research works are still going on in this topic. We have varied it from as low as 1 min upto 3 hours, since the upper range is suggested so according to a study \cite{van2020aerosol} in the New England Journal of Medicine. However, upon availability of more legitimate information about spatial and temporal thresholds, our algorithm should work just fine with the updated parameter values, without any modification in its design.

\subsection{Experiment with \emph{BD Cellphone}}
\begin{figure}
\centering
\vspace{-25pt}
\subfloat[]{\includegraphics[trim = 20mm 70mm 10mm 60mm, clip, width=0.300\textwidth]{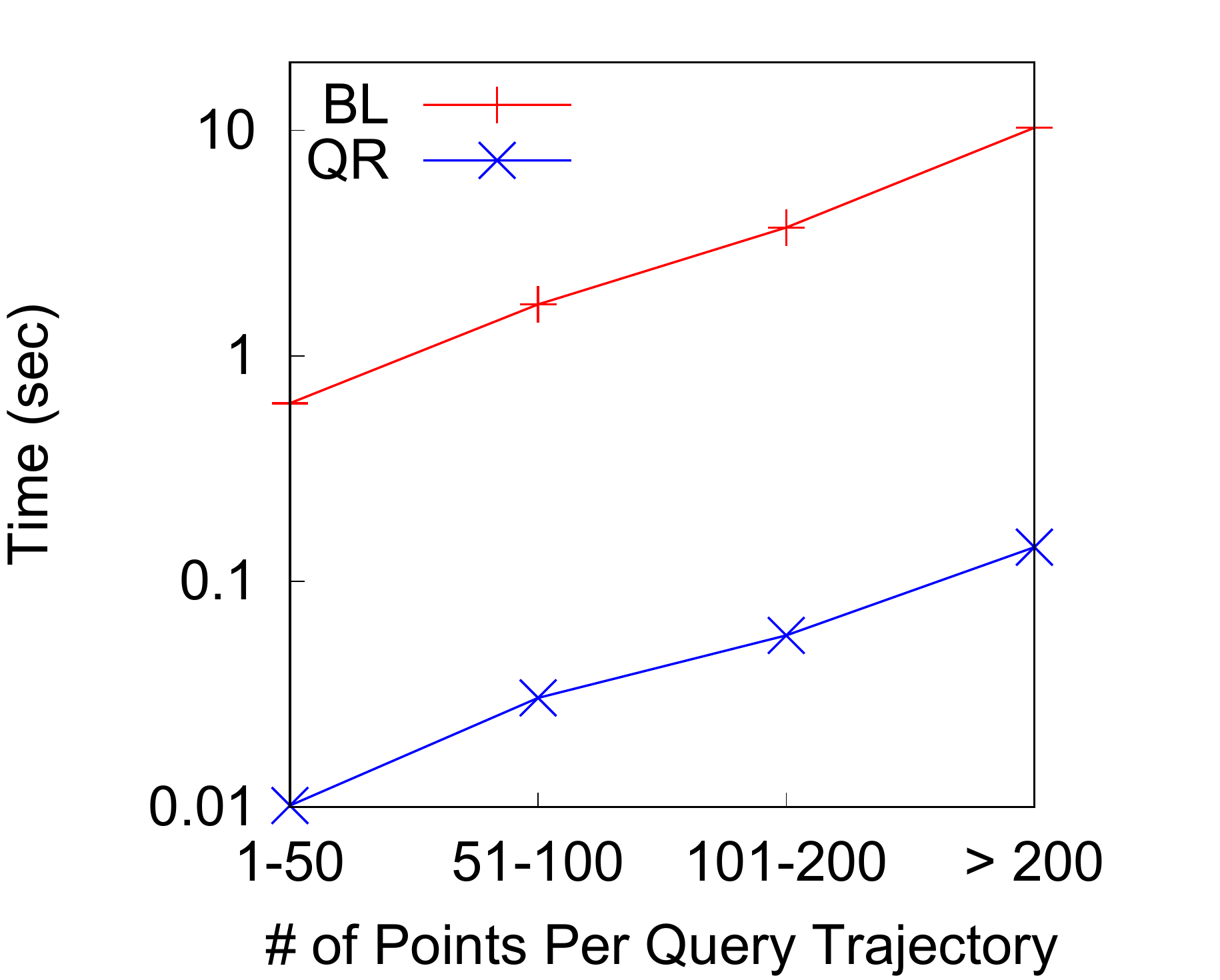}\label{fig:point_t}}
\subfloat[]{\includegraphics[trim = 40mm 70mm 0mm 60mm, clip, width=0.280\textwidth]{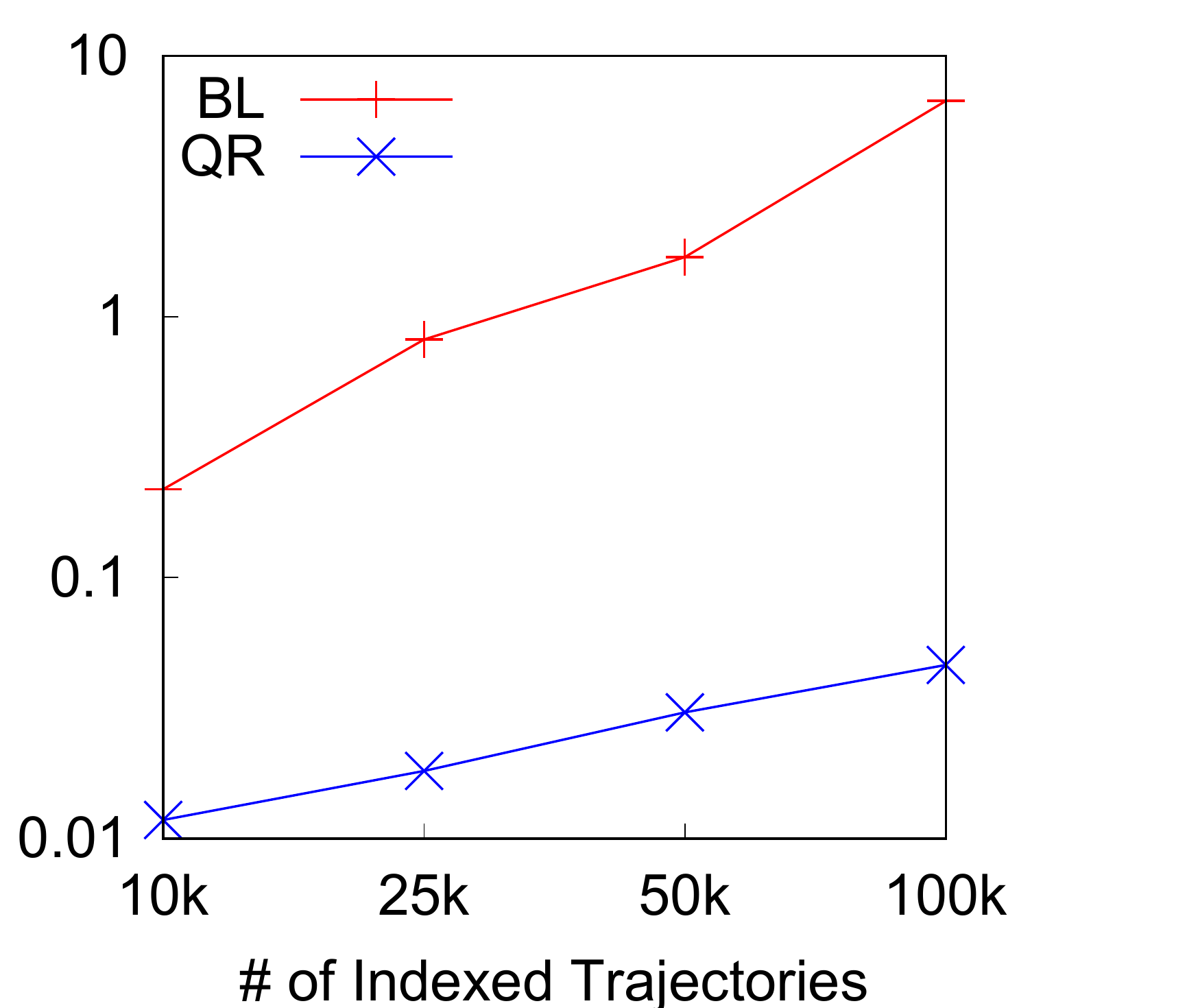}\label{fig:traj_t}}
\vspace{-10pt}
\subfloat[]{\includegraphics[trim = 10mm 60mm 10mm 60mm, clip, width=0.295\textwidth]{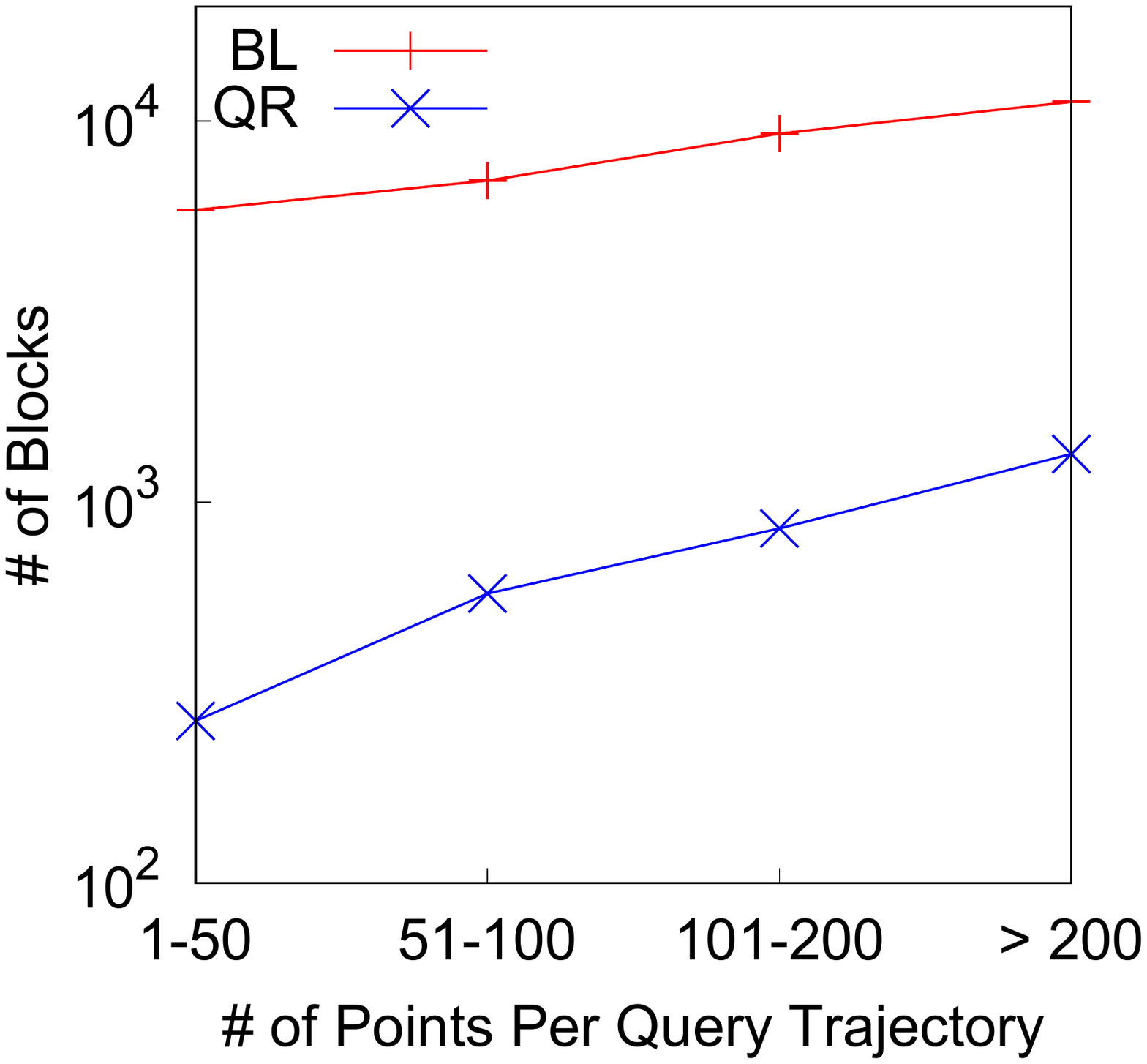}\label{fig:point_io}}
\subfloat[]{\includegraphics[trim = 30mm 60mm 0mm 60mm, clip, width=0.277\textwidth]{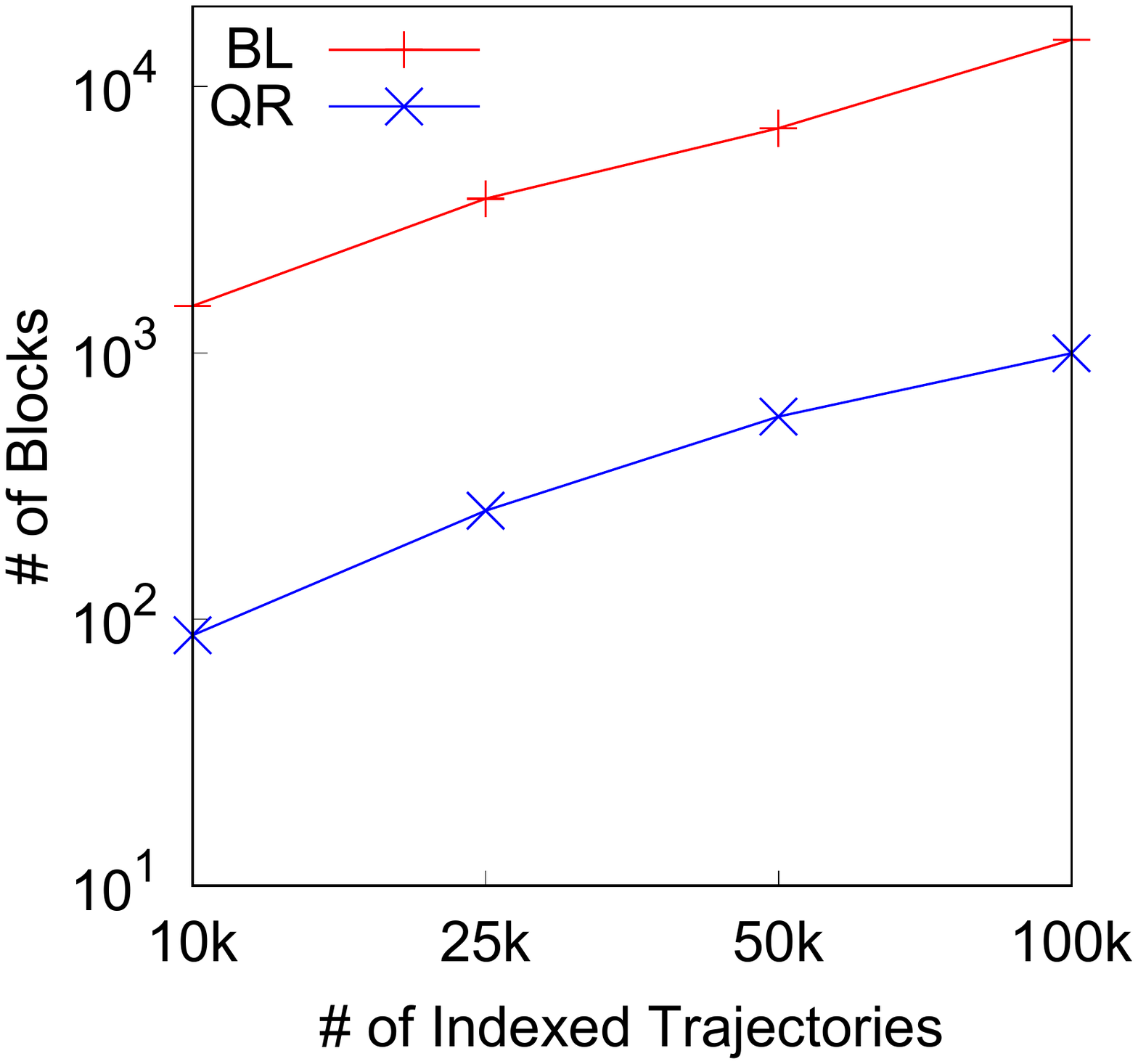}\label{fig:traj_io}}
\vspace{-10pt}
\caption{Evaluating {\ctq} for varying no. of points per query trajectory (a \& c)  and no. of trajectories (b \& d)}
\label{fig:ctq_point_traj}
\end{figure}

\begin{figure}
\centering
\vspace{-25pt}
\subfloat[]{\includegraphics[trim = 20mm 70mm 10mm 60mm, clip, width=0.300\textwidth]{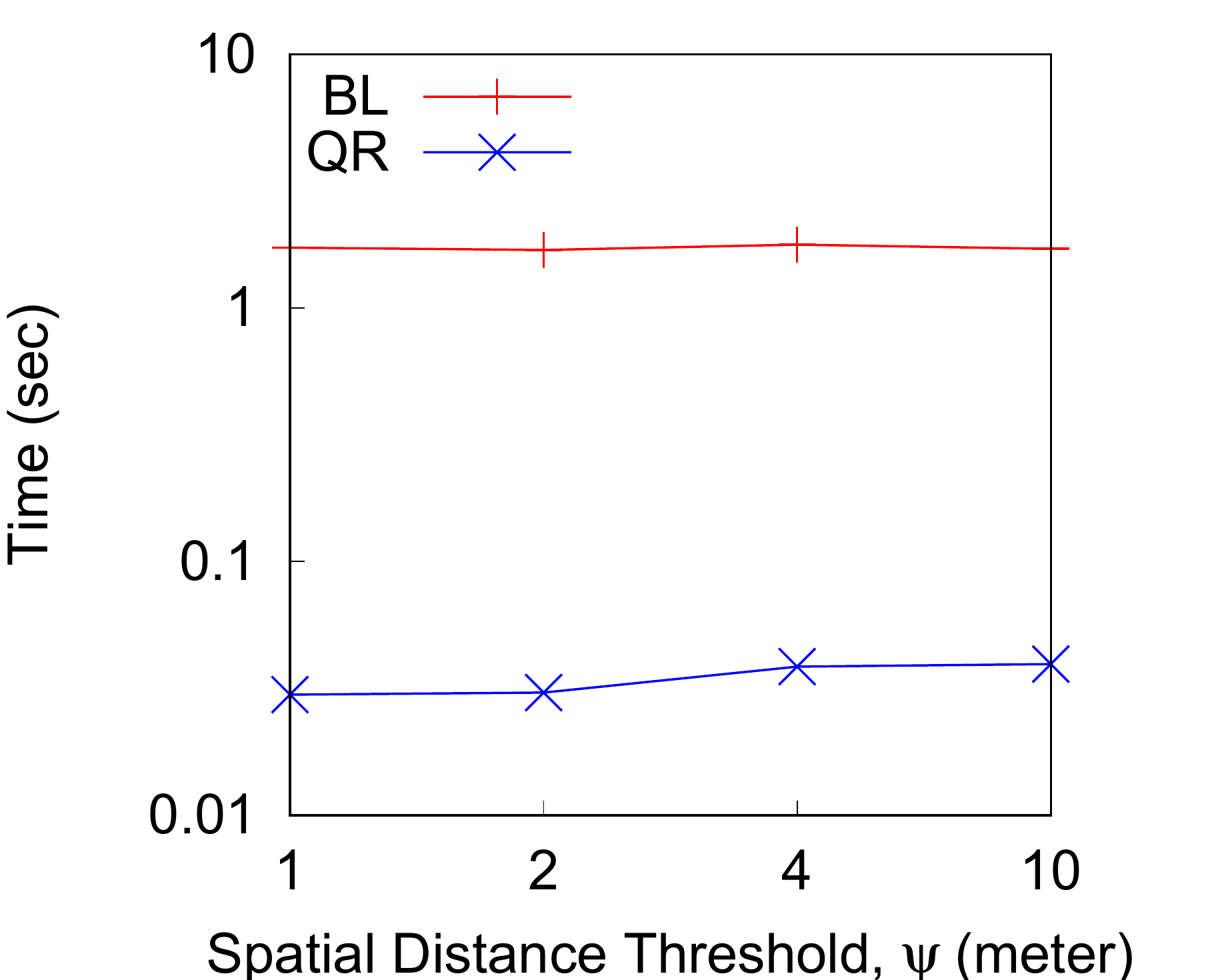}\label{fig:psi_t}}
\subfloat[]{\includegraphics[trim = 40mm 70mm 0mm 60mm, clip, width=0.280\textwidth]{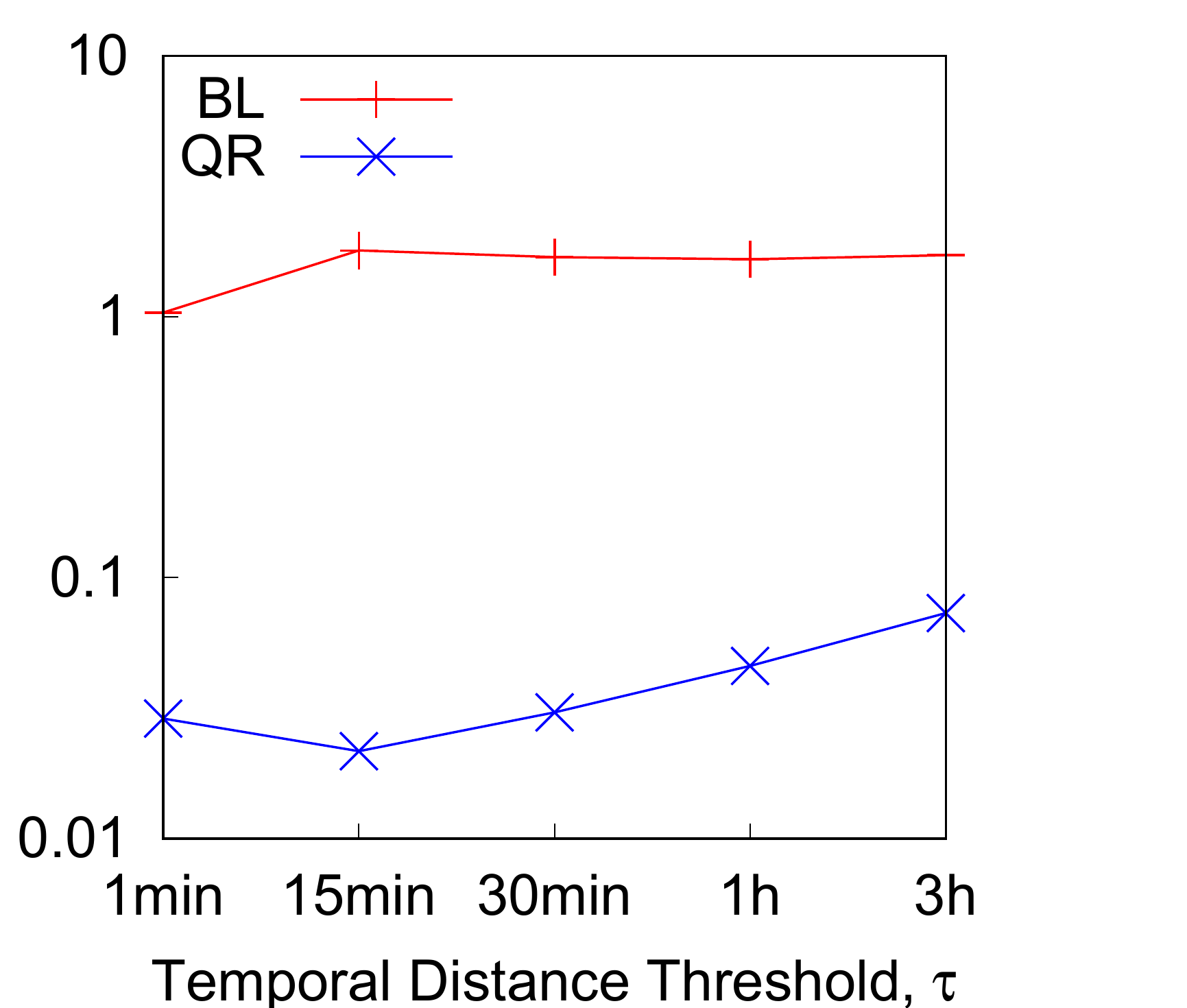}\label{fig:tau_t}}
\vspace{-10pt}
\subfloat[]{\includegraphics[trim = 10mm 60mm 10mm 60mm, clip, width=0.295\textwidth]{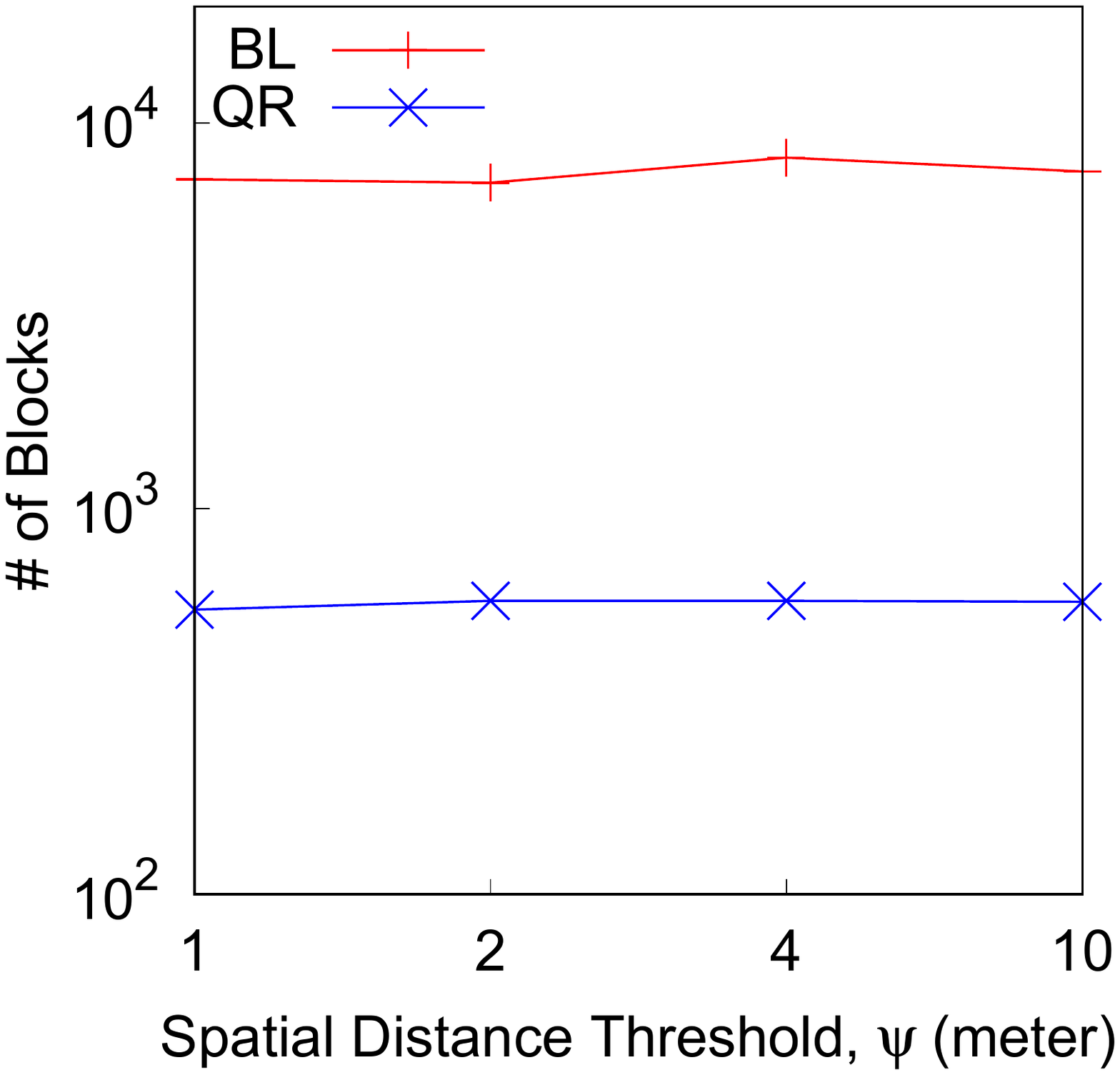}\label{fig:psi_io}}
\subfloat[]{\includegraphics[trim = 30mm 60mm 0mm 60mm, clip, width=0.279\textwidth]{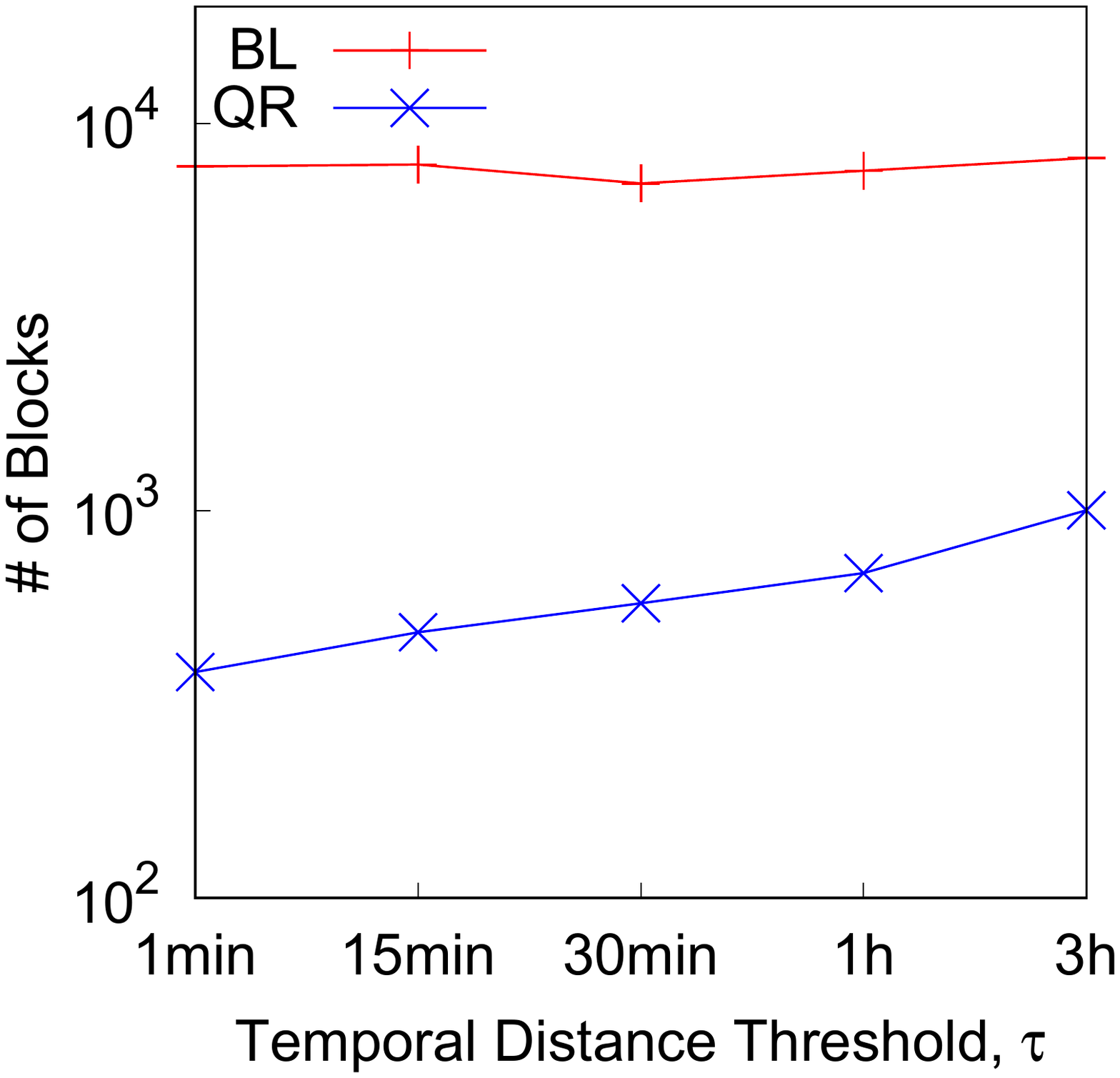}\label{fig:tau_io}}
\vspace{-10pt}
\caption{Evaluating {\ctq} for varying spatial distance threshold, $\psi$ (a \& c)  and temporal distance threshold, $\tau$ (b \& d)}
\label{fig:ctq_psi_tau}
\end{figure}

\begin{figure}
\centering
\vspace{-25pt}
\subfloat[]{\includegraphics[trim = 20mm 60mm 30mm 60mm, clip, width=0.250\textwidth]{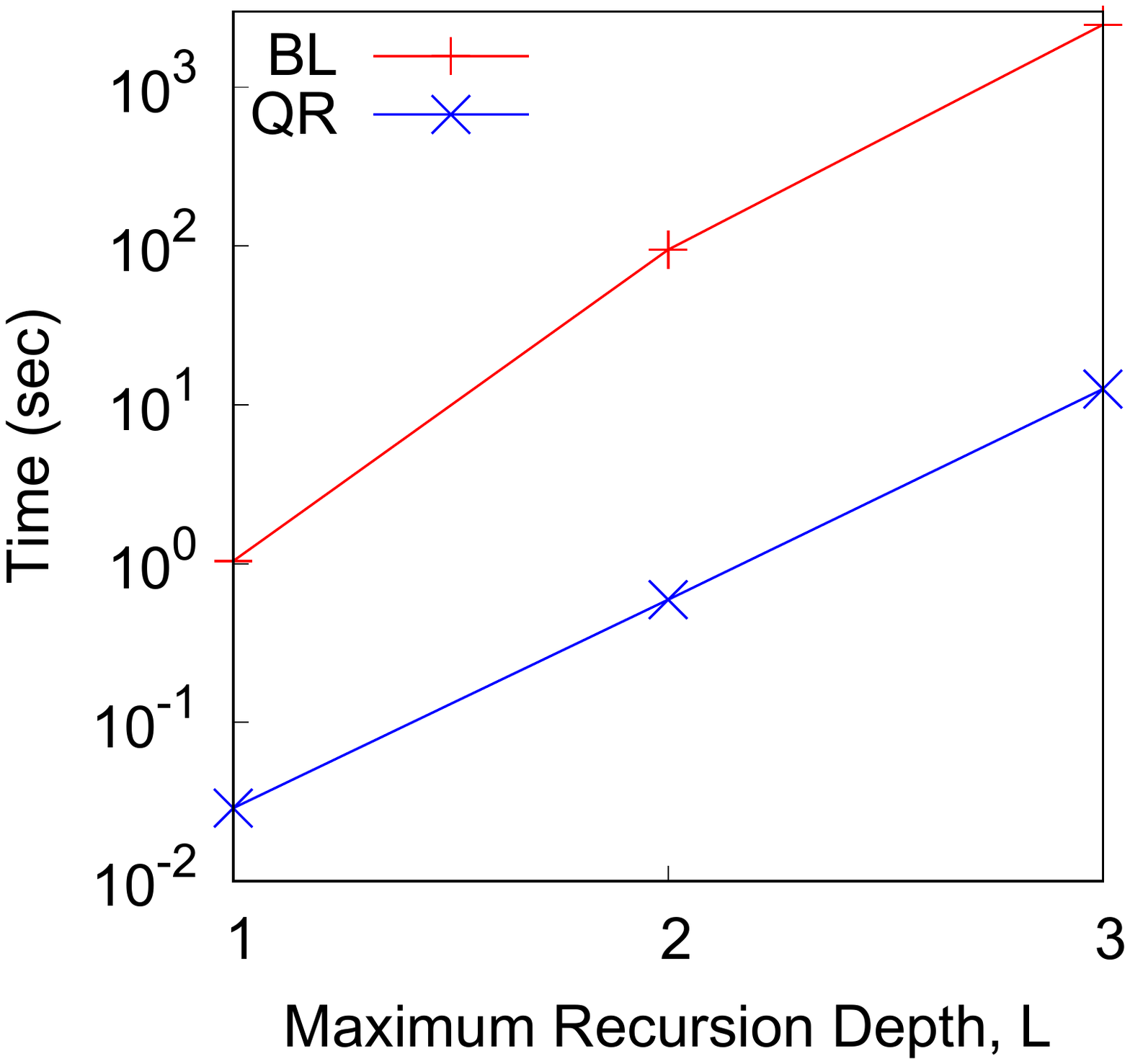}\label{fig:L_t}}
\subfloat[]{\includegraphics[trim = 15mm 60mm 30mm 60mm, clip, width=0.252\textwidth]{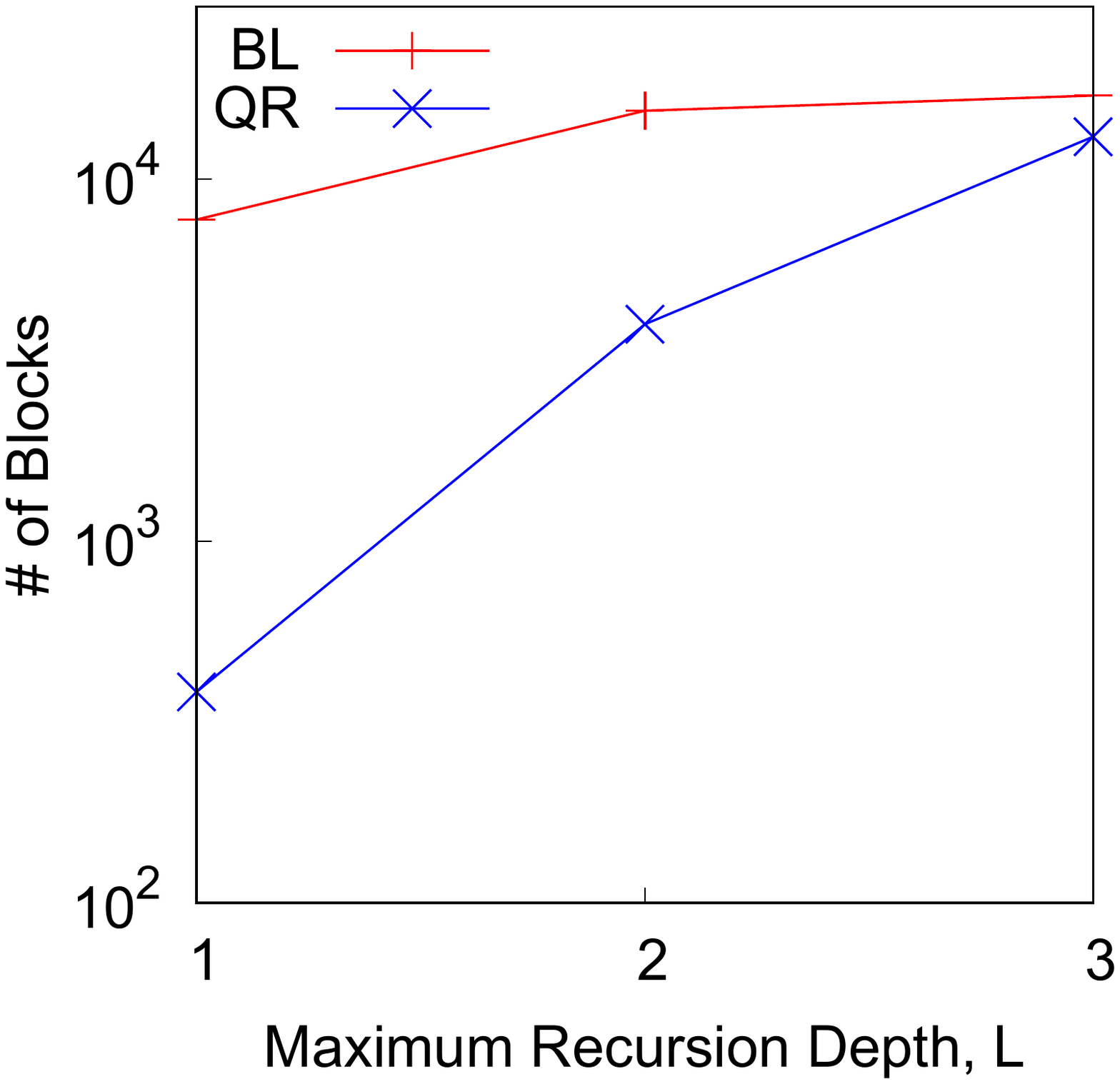}\label{fig:L_io}}
\vspace{-10pt}
\caption{Evaluating {\ctq} for varying maximum recursion depth, $L$}
\label{fig:ctq_L}
\end{figure}

\noindent{\bf \emph{(i) No. of points per query trajectory: }}Our algorithm using the {\qr} index outperforms the baseline by 
$2$ orders of magnitude in terms of both runtime (Figure~\ref{fig:point_t}) and by $1-2$ orders of magnitude in terms of I/O cost (Figure~\ref{fig:point_io}).
As the number of points in query trajectory increases, more user trajectories at different blocks are expected to be processed. So runtime as well as I/O should increase in both the approaches, as reflected in the graph (Figure~\ref{fig:point_t}, ~\ref{fig:point_io}).

\noindent{\bf \emph{(ii) No.\ of indexed trajectories: }} The {\qr} outperforms the baseline by around $2$ orders of magnitude in terms of runtime (Figure ~\ref{fig:traj_t}) and by around $1$ order of magnitude (Figure ~\ref{fig:traj_io}) in terms of disk I/O cost. Both the approaches follow an increasing trend with the increase in the number of indexed trajectories. This is because more trajectories require more disk blocks to be stored. So a higher number of disk blocks, i.e., larger number of trajectories, are expected to be retrieved in query processing, requiring more time to be processed.

\noindent{\bf \emph{(iii) Spatio-temporal thresholds ($\psi$, $\tau$): }} When we vary spatial ($\psi$) or temporal distance thresholds ($\tau$), {\qr} works better than the baseline by at around $1-2$ orders of magnitude in terms of runtime (Figure ~\ref{fig:psi_t}, ~\ref{fig:tau_t}) and by at least $1$ order of magnitude (Figure ~\ref{fig:psi_io}, ~\ref{fig:tau_io}) in terms of disk I/O cost. 

\noindent{\bf \emph{(iv) Maximum recursion depth ($L$): }} The performance of the {\qr} is significantly important when we consider multiple levels of {\ctq}, that is, when we consider exposure from already exposed users upto a certain level, instead of confirmed patients only. The {\qr} outperforms baseline approach by $2-3$ orders of magnitude in terms of runtime (Figure ~\ref{fig:L_t}) and by around $1-2$ orders of magnitude in terms of I/O cost (Figure ~\ref{fig:L_io}) when we vary maximum recursion depth level, $L$. More importantly, note that, the {\qr} can provide results in tens of seconds in case of upto three levels of exposure while the baseline would require thousands of seconds to do that. The benefits of I/O may seem misleading for higher depth levels (Figure ~\ref{fig:L_io}) because the {\ctq} processing gets saturated in terms of disk block access, i.e. it accesses almost all the blocks in both approaches (baseline being marginally higher) to retrieve potential candidate trajectories. For this reason, the baseline approach has a somewhat flat tail for already accessing all the disk blocks. But running the experiment with higher number of trajectories to demonstrate this I/O gain is not feasible because of the intractable runtime of the baseline method. Besides, instead of default temporal distance threshold ($\tau$) value of 15 minutes, we have used $\tau$ = 1 minute for running this experiment to keep the results demonstrable.

\subsection{Experiments with \emph{NYF}}
\begin{figure}
\centering
\vspace{-25pt}
\subfloat[]{\includegraphics[trim = 20mm 70mm 20mm 60mm, clip, width=0.288\textwidth]{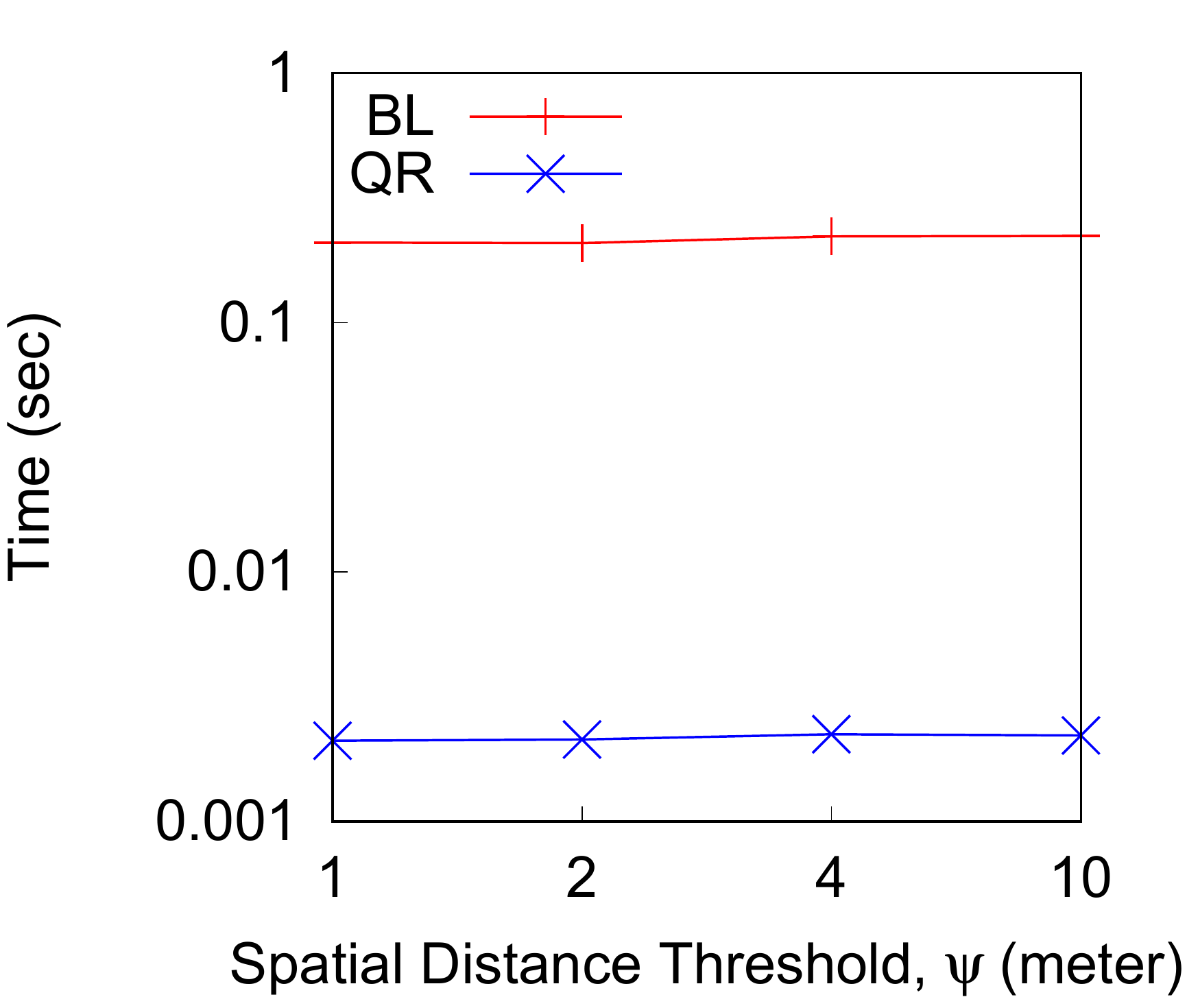}\label{fig:nyf_psi_t}}
\subfloat[]{\includegraphics[trim = 42mm 70mm 10mm 60mm, clip, width=0.270\textwidth]{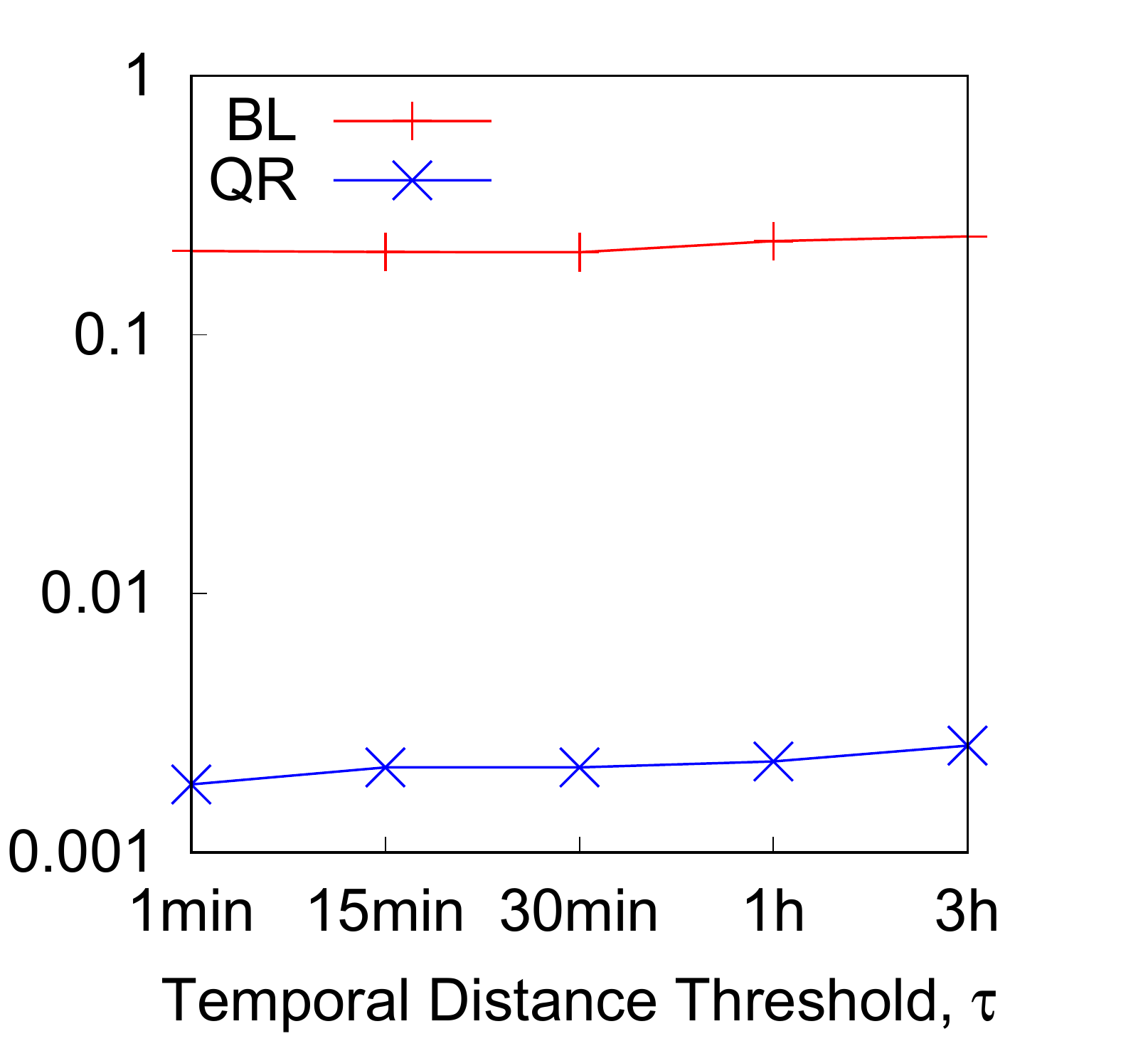}\label{fig:nyf_tau_t}}
\vspace{-10pt}
\subfloat[]{\includegraphics[trim = 20mm 70mm 20mm 60mm, clip, width=0.288\textwidth]{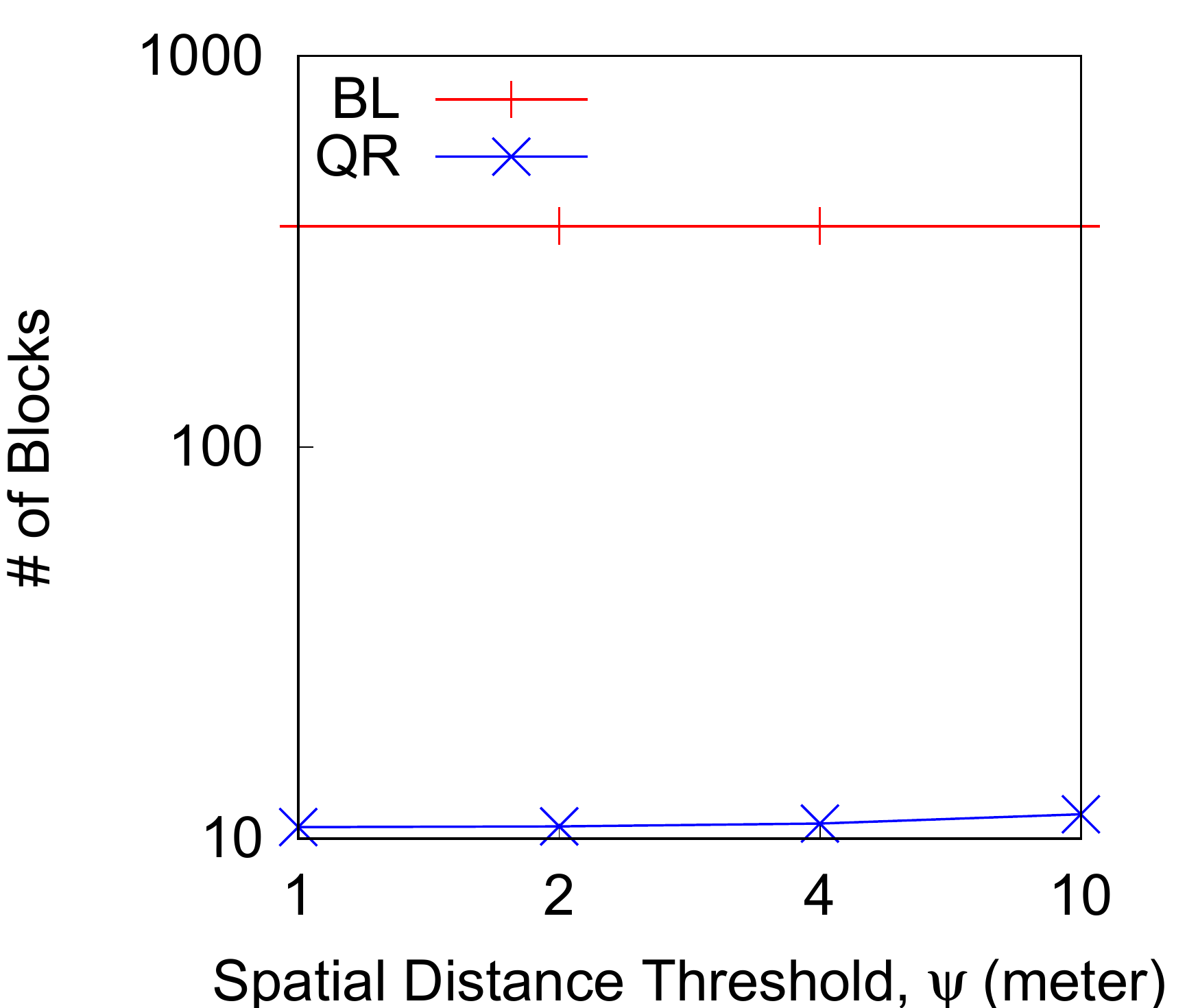}\label{fig:nyf_psi_io}}
\subfloat[]{\includegraphics[trim = 42mm 70mm 10mm 60mm, clip, width=0.270\textwidth]{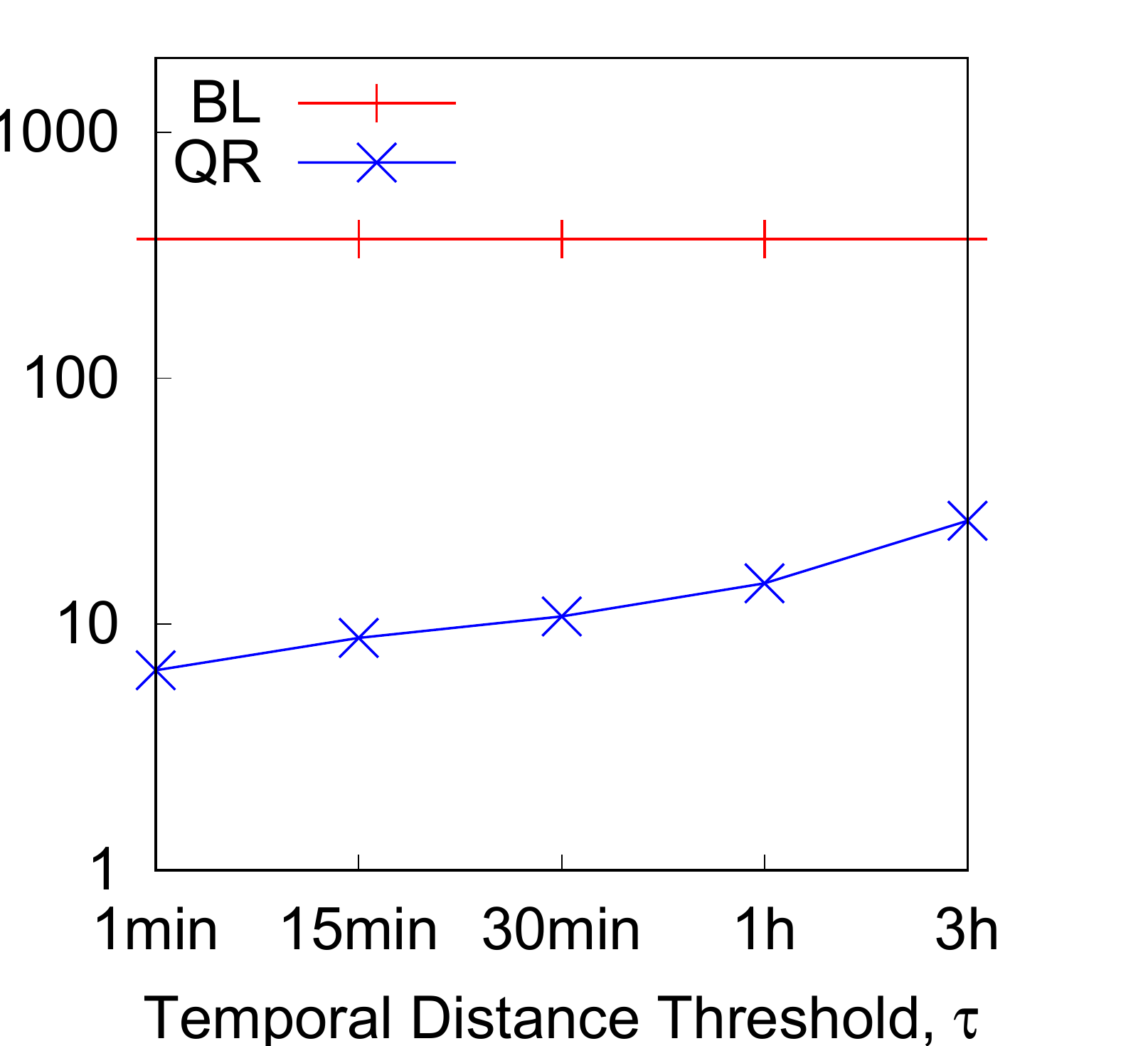}\label{fig:nyf_tau_io}}
\vspace{-10pt}
\caption{Evaluating {\ctq} on NYF for varying spatial distance threshold, $\psi$ (a \& c)  and temporal distance threshold, $\tau$ (b \& d)}
\label{fig:ctq_nyf_psi_tau}
\end{figure}

The \emph{NYF} dataset is significantly smaller than the \emph{BD Cellphone} dataset. We report only the impacts of varying spatio-temporal distance thresholds in the experiments with this dataset since varying the other parameters would be of little value for the dataset size.

\noindent{\bf \emph{(i) Spatio-temporal thresholds ($\psi$, $\tau$): }} The {\qr} works better than the baseline by at around $2$ orders of magnitude in terms of runtime (Figure ~\ref{fig:nyf_psi_t}, ~\ref{fig:nyf_tau_t}) and by around $1-2$ orders of magnitude (Figure ~\ref{fig:nyf_psi_io}, ~\ref{fig:nyf_tau_io}) in terms of I/O cost. So the performance of {\qr} is comparatively even better for \emph{NYF} dataset. The I/O graph of baseline looks flat because all the disk blocks have been accessed by it whatever the parameter values are. This is because the dataset spans over a longer temporal domain than that of \emph{BD Cellphone}.

\subsection{Evaluation of \qqr}
\begin{figure}[!h]
\centering
\includegraphics[trim = 20mm 60mm 30mm 60mm, clip, width=0.280\textwidth]{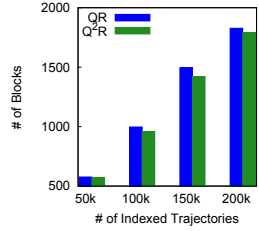}
\vspace{-10pt}
\caption{I/O comparison of enhanced tree for {\ctq}}
\label{fig:qrVSqqr_io}
\end{figure}

In a real system, the performance gain by our proposed index will largely be attributed to the lower I/O cost. This is not simulated in the runtime experiments. So the merits demonstrated for the {\qr} is very likely to be manifold to what is reported in case of a real deployment.

The further enhancement we have proposed on the {\qr}, namely the {\qqr} works slightly better in terms of I/O cost specially when we deal with larger number of trajectories, as demonstrated in Figure ~\ref{fig:qrVSqqr_io}. Note that, this experiment is run by indexing 50k, 100k, 150k and 200k trajectories respectively. The {\qqr} achieves $2-5\%$ reduction in the number of disk blocks accessed, specially for higher number of trajectories. So, though the {\ctq} processing using the {\qqr} needs multiple levels of tree traversal, marginally lower I/O overhead can eventually result in better runtime performance as well, which is subject to further experiments in real systems.
\section{Related Works}
\label{related}
The works related to ours mostly encompass studies in trajectory indexing in spatio-temporal domain and some query processing using these indexes. Besides there are many ongoing researches in contact tracing with the outbreak of COVID-19 pandemic, most of which attempt to solve the challenge from different perspectives.

\subsection{Spatio-temporal Trajectory Indexing}
Indexing of moving objects i.e. storing trajectory data efficiently has received considerable attention throughout the last two decades. Mokbel et al. present a summary of spatio-temporal indexing methods in their survey \cite{mokbel2003spatio} according to some of the earlier studies in this field. They point out three techniques that have been used to index historic trajectory data. These are augmentation of temporal index with existing spatial index, combining both spatial and temporal access in a single structure and indexing mainly based on temporal information while treating spatial index as secondary. Nguyen-Dinh et al. extend the work in \cite{nguyen2010spatio} and summarize the indexing methods adopted in 2003-2010 period according to the aforementioned categories. Mahmood et al. focus on the more recent techniques in \cite{mahmood2019spatio}, in succession of the previous works. We describe some of these indexing methodologies briefly and present comparative arguments of the relevant ones with our work. The details and more elaborate discussion can be found in \cite{mokbel2003spatio}, \cite{nguyen2010spatio}, \cite{mahmood2019spatio} and the research works they have addressed in these studies.

RT-tree, 3D R-tree etc. indexing methods deal with temporal information along with spatial data as summarized in \cite{mokbel2003spatio}. RT-tree \cite{xu1990improved} simply augments the time interval information with the MBRs of R-tree. So it achieves a performance as good as R-tree for spatial queries but the temporal queries often span the whole tree. Our proposed contact tracing query need to process both spatial and temporal ranges, so it would be inefficient for our purpose. On the other hand, 3D R-tree \cite{theodoridis1996spatio} considers temporal attribute as an additional dimension with the spatial R-tree, processing spatial and temporal queries alike. We have used this approach as our baseline method for its potential applicability to our proposed query. Spatio-temporal R-tree (STR-tree) \cite{pfoser2000novel} is another approach to index spatio-temporal data with R-tree at the core but with different insertion and splitting strategy. It focuses on both spatial locality and trajectory preservation based on a configurable parameter \cite{mokbel2003spatio}. However different segments of a trajectory may be stored in different nodes or spatio-temporally close trajectories may be grouped separately in this approach, both of which can deteriorate the performance of our proposed query processing.

Some trajectory oriented access methods puts more emphasis on grouping the points of each trajectory together. TB-tree \cite{pfoser2000novel}, SETI \cite{chakka2003indexing} etc. adopt such mechanisms. Spatial queries and keeping spatially closed objects together are not among their primary concerns as stated in \cite{mokbel2003spatio}.

Trajectory-bundle tree (TB-tree) strictly emphasizes on trajectory preservation and gives up on spatial locality if needed. It is also built on top of R-tree, which means, the MBRs of TB-tree overlap a lot in contrary to its minimization as would be done in regular R-tree. This structure can deal with trajectory based queries involving topology with spatio-temporal attributes like area, time etc. or those based on navigation quite efficiently. But in the contact tracing query we need both trajectory interaction in terms of spatio-temporal locality and trajectory preservation for its efficient processing. So it not readily applicable to our problem as well. Besides, both TB-tree and STR-tree retrieves trajectory segments incrementally \cite{pfoser2000novel}. If someone is infected in our case, her whole trajectory needs to be retrieved, which would be costly using these indexes.

The indexing mechanism SETI, proposed in \cite{chakka2003indexing} addresses the scalability issues of existing indexing schemes. It presents a two level index structure: the first level index partitions the spatial domain into static, uniform and non-overlapping cells, the second level index uses a traditional R-tree to index the time domain. Using the first index, the segments of each trajectory are assigned to the cells, according to their spatial coordinates. If a segment spans multiple cells then it is split at the cell boundary. Then, the time span of the segments (i.e the minimum and the maximum timestamp) in each cell are saved in an R-tree. So, effectively the spatial and temporal dimensions are decoupled in the process. The authors mainly discussed range queries with their proposed index where a spatial and then a temporal filtering is done using the first and second level index respectively. After that, a refinement step retrieves the desired trajectories. The index also supports efficient insert, delete and update.

The Start-End timestamp B-tree (SEB-tree) \cite{song2003seb} is another trajectory oriented indexing similar to SETI. The space is partitioned into overlapping zones which are indexed using SEB-tree considering only start and end timestamps. The moving objects are mapped to their zones using hashing. But unlike SETI, SEB-tree works on the two dimensional points instead of the trajectories.

Some of the other indexing schemes \cite{mokbel2003spatio} presents for indexing historic trajectory data include MR-tree, HR-tree, MV3R-tree etc. These indexes disintegrate spatial and temporal dimensions as they aim at storing spatial attributes of the trajectories at different timestamps in different R-trees.

\cite{nguyen2010spatio} presents some specialized indexing methods besides some improvements on the previous works. The MTSB-tree \cite{zhou2005close} has its similarity with SETI in terms of spatio-temporal organization with the difference that it uses Time-Split B-tree (TSB-tree) for temporal indexing of trajectory segments instead of R-tree based temporal index of SETI. So trajectory segments are sorted in increasing order of time. FNR-tree and MON-tree use multiple R-trees to store object movement locations and time intervals. The latter one also uses a hash structure for mapping object movement lines to the lower level temporal R-tree. GS-tree indexes trajectories in a constrained graph by dividing them into nodes and edges. It is a balanced binary tree that discriminates time dimension from the spatial counterpart. Here a leaf node represents MBR of edges and points to two different data structures for spatial and temporal dimensions. The Compressive Start-End tree (CSE-tree) \cite{wang2008flexible} divides space in disjoint regions like SETI and maintains temporal indexes for each of these regions. It considers time intervals as two dimensional points and maintains separate B+ trees for indexing end times followed by start times to group trajectory segments. Polar tree is specialized to index direction of the moving objects. It uses an in-memory unbalanced binary tree to index orientation of objects with respect to a given focal point. It can efficiently determine if many objects get close to or far from a reference site. Besides, RTR-tree and TP$^2$R-tree provides better support for range queries in euclidean space in indoor environments. The other indexing methods studied in the survey are out of scope of this literature as they bear little resemblance to our work.

\cite{mahmood2019spatio} describes some of the more recent works in trajectory indexing. TrajStore partitions trajectories and clusters spatio-temporally close segments together on the disk. TrajTree also relies on trajectory segmentation where leaf nodes contain sub-trajectories and non-leaves hold sequences of the bounding boxes of their child nodes. Most of the other spatio-temporal indexes are specialized and application specific. For instance, UTH and UT$_{GRID}$ deals with trajectories with uncertain portions, PARINET is specific to trajectories along road networks, TRIFL optimizes trajectory indexing in flash storage and so on.

\subsection{State of the Art Queries}
The common queries in spatio-temporal domain can be classified into two broad categories, coordinate based query and trajectory based query \cite{pfoser2000novel}. Coordinate based queries include point spefic query, range query, nearest neighbor search etc. while trajectory based query can involves topology or navigational details. Coordinate based queries have been addressed since the earlier works in the spatio-temporal trajectory domain like RT-tree, 3D R-tree \cite{mokbel2003spatio} etc., which were improved later in more efficient indexes like SETI, SEB-tree etc. Trajectory oriented queries where trajectory preservation can play an important role was addressed and efficiently processed using STR-tree, TB-tree in \cite{pfoser2000novel}. The later indexes proposes improvements in both directions, works with new queries like $k$-NN (e.g. TrajTree), but these queries do not align with the proposed novel contact tracing query since both spatio-temporal range search and whole trajectory retrieval are of utmost importance here.

A query somewhat similar to the contact tracing query (CTQ) is presented in \cite{Yadamjav2018EfficientMQ}, which the authors call trajectory multi-range query (MRQ). The goal of MRQ is to find the set of trajectories that go through a set of given spatio-temporal ranges. We can consider CTQ as a multi-range query over trajectories too, where there is a query spatio-temporal range for each point of the CTQ query trajectory. However, there is a very important distinction between MRQ and CTQ. In MRQ, the resultant trajectories pass through all the given spatio-temporal ranges, whereas, in the case of CTQ, even if a trajectory goes through only one of the query spatio-temoral ranges, we need to return it and do further processing on it. Also, in CTQ we consider indirect contact/exposure to the query trajectory which is not considered in MRQ.

\subsection{Recent Progress in Contact Tracing}
Since the start of the COVID-19 pandemic, governments have rolled out contact tracing apps (\cite{10.1215/18752160-8698301}) in order to contain the spread of the virus. The aim of these apps is to understand if a user was exposed to a known COVID infected person, and if so, notify her for testing and starting the quarantine process. A review of the existing technologies for contact tracing is presented in \cite{legendre2020contact}. Proximity based contact tracing apps use Bluetooth and WiFi to infer relative proximity to other users. Location based technologies, on the other hand, use GPS to locate the exact position of the users. Although GPS positioning is not very accurate in indoor spaces, that problem can be overcome with the additional use of crowd-sourced WiFi localisation. With modern WiFi access points the accuracy is good enough for contact tracing \cite{legendre2020contact}. However, both of these two technologies require a smartphone and an app to be installed by the user. The infrastructure and devices required by these methods may not be available in certain places, specially in developing countries. In such situations, using mobile operator's infrastructure to locate the phone of a user is an option. This has the advantage of not requiring the user to do anything and the usage of existing infrastructure. However, there still remains the accuracy and privacy concerns.
\section{Conclusions}
\label{conclusion}
We have proposed a novel {\ctq} in the context of spatio-temporal databases and developed a multi-level index, namely {\qr}, to efficiently process the {\ctq}. Experimental results show that the {\qr} based approach outperform the baseline by 1-2 orders of magnitude both in terms of processing time and I/O.
In future, we plan to develop a system based on the proposed index and make it available for the community.

\bibliography{CT} 

\end{document}